\documentclass[a4paper,11pt]{article}
\pdfoutput=1 

\usepackage{jheppub} 

\usepackage[T1]{fontenc} 
\usepackage{appendix}

\usepackage{amssymb,amsfonts}
\usepackage{amsmath}
\usepackage{mathtools}
\usepackage{amsthm}
\usepackage{empheq}
\usepackage[colorinlistoftodos]{todonotes}
\usepackage{array}

\renewcommand{\arraystretch}{1.2}

\newcommand{\be}{\begin{equation}}
\newcommand{\ee}{\end{equation}}
\newcommand{\bea}{\begin{eqnarray}}
\newcommand{\eea}{\end{eqnarray}}

\def\Dslash{\,\,{\raise.15ex\hbox{/}\mkern-13mu D}}
\def\Dbarslash{\,\,{\raise.15ex\hbox{/}\mkern-12mu {\bar D}}}
\def\delslash{\,\,{\raise.15ex\hbox{/}\mkern-10mu \partial}}
\def\delbarslash{\,\,{\raise.15ex\hbox{/}\mkern-9mu {\bar\partial}}}
\def\pslash{\,\,{\raise.15ex\hbox{/}\mkern-11mu p}}
\def\qslash{\,\,{\raise.15ex\hbox{/}\mkern-9mu q}}
\def\kslash{\,\,{\raise.15ex\hbox{/}\mkern-11mu k}}
\def\eslash{\,\,{\raise.15ex\hbox{/}\mkern-9mu \epsilon}}

\newcommand{\slsh}[1]{\,\,{\raise.15ex\hbox{/}\mkern-12mu {#1}}}

\definecolor{myorange}{RGB}{199,146,32}

\setcitestyle{square}

\title{\boldmath Dynamical Systems and Superstring Phases in the Early Universe }

\author[a]{Noelia S\a'anchez Gonz\a'alez,}
\author[a]{Joseph P. Conlon,}
\author[b]{Edmund J. Copeland,}
\author[a]{Edward Hardy}

\affiliation[a]{Rudolf Peierls Centre for Theoretical Physics,\\Beecroft Building,\\ Parks Rd, Oxford, OX1 3PU,\\UK}
\affiliation[b]{School of Physics and Astronomy, \\
University of Nottingham, \\
Nottingham, \\
NG7 2RD, UK
}

\emailAdd{noelia.sanchezgonzalez@physics.ox.ac.uk}
\emailAdd{joseph.conlon@physics.ox.ac.uk}
\emailAdd{ed.copeland@nottingham.ac.uk}
\emailAdd{edward.hardy@physics.ox.ac.uk}

\preprint{May 2025}

\abstract{We study the string theory dynamics of the volume scalar rolling down an exponential potential during the period between inflation and reheating, in a background of cosmic superstring loops. In the context of the LVS potential, we demonstrate the existence of a novel string loop attractor tracker solution, in which 75\% of the energy density of the universe is in the form of a gas of fundamental cosmic superstring loops (a configuration preferred over the standard radiation tracker). On this tracker, it is the continual reduction in the string tension as the volume scalar evolves that makes the loops stable against decay. For more general non-LVS potentials, mixed radiation-loop trackers can also occur.

}

\begin{document} 
\maketitle
\flushbottom

\section{Introduction and Review}\label{Sec1}

Early universe cosmology represents one of the promising locations to look for 
phenomenological effects of string theory (for reviews of string cosmology, see \cite{Cicoli:2023opf, Brandenberger:2023ver}). The ambient energies are high and much of the history, in particular that between inflation and nucleosynthesis, is currently observationally unconstrained but might be probed by future searches e.g. for gravitational waves.
\\
\\
Recently, there has been much exploration of scenarios where moduli fields traverse large distances in field space during the early universe \cite{Conlon:2022pnx, Revello:2023hro, Shiu:2023fhb, Shiu:2023nph, Seo:2024qzf, Apers:2024ffe, Shiu:2024sbe, Andriot:2024jsh, Conlon:2024uob, Revello:2024gwa, Cicoli:2024yqh, Casas:2024oak, Apers:2024dtn, Brunelli:2025ems, Andriot:2025cyi, Ghoshal:2025tlk} (see \cite{Conlon:2008cj, Cicoli:2015wja} for some older papers), in particular in the period between the inflationary epoch and nucleosynthesis. Such explorations are motivated by the notion that the present vacuum may live near the boundary of moduli space -- large hierarchies are required in the present universe, and asymptotic vacua represent one of the most natural ways to generate such hierarchies. In this case, fields must traverse large distances in the early universe as the system evolves from the centre of moduli space towards the present vacuum in the asymptotic regime.
\\
\\
Although string compactifications contain large numbers of moduli, the vast majority are `internal' moduli, whose vevs do not lead towards the asymptotic regions of moduli space. The two moduli whose vevs do most naturally lead to asymptopia are the dilaton and volume modulus. However, as the dilaton controls the string coupling ($\textrm{Re} (S) = g_s^{-1}$ in type IIB), it is very hard to obtain phenomenologically viable vacua with large dilaton vevs. With $g_s$ extremely small, it is next-to-impossible to incorporate Standard Model gauge sectors with $\alpha_{SM}^{-1} \sim 30$. This leaves the volume modulus, $\mathcal{V}$, as the one field that can traverse large trans-Planckian distances in field space, rolling deep into the asymptotic regions of moduli space while still remaining entirely compatible with semi-realistic final vacua (through local Standard Model constructions such as \cite{Aldazabal:2000sa, Berenstein:2001nk, Verlinde:2005jr, Beasley:2008dc, Conlon:2008wa, Krippendorf:2010hj, Cicoli:2021dhg} (see \cite{Quevedo:2014xia, Marchesano:2024gul} for reviews) and LVS-style moduli stabilisation scenarios \cite{Balasubramanian:2005zx, Conlon:2005ki}).
\\
\\
It is natural to expect that the cosmological evolution while the modulus is rolling begins as kination due to the steep exponential potentials expected in string theory (see \cite{Brustein:1992nk} for an early discussion), with the energy density of the universe dominated by the kinetic energy of the modulus. Given sufficient time, the dynamics can subsequently approach a tracker solution, where the fraction of energy in different components of the energy density becomes time-independent. Extended kination or tracker cosmological solutions involve (almost by definition, see Eq. \ref{kinevolution}) significant trans-Planckian field excursions, making them intrinsically interesting from a string theory perspective.
Recent interest in extended volume modulus kination/tracker solutions in string theory has led to a plethora of results \cite{Apers:2024ffe, Conlon:2024uob, Revello:2024gwa, Brunelli:2025ems, Ghoshal:2025tlk}. Our focus here is on the novel behaviour of macroscopically large fundamental strings during an epoch of large displacements in the compactification volume (which includes tracker epochs as well as kination, in both of which the volume modulus evolves significantly). 
\\
\\
The study of fundamental strings in cosmology is interesting as
cosmic superstring networks are one of the few ways actual fundamental strings could exist today in the present universe \cite{Copeland:2003bj} (for reviews see \cite{Vilenkin:2000jqa, Copeland:2009ga}). Strings are characterised by their tension: in string theory, all parameters depend on the vevs of moduli, and so the string tension (more precisely, the ratio of the string tension to the 4d Planck scale) is set by the moduli vevs. The string scale $m_s$ is directly sensitive to the volume (as $m_s \sim g_s M_P/\sqrt{\mathcal{V}})$. When the volume modulus rolls, all physical scales -- including the string tension -- also change. 
\\
\\
This implies that, during volume-driven kination or tracker epochs, the dynamics of cosmic superstrings is the dynamics of strings with time-varying tension. 
It was shown in \cite{Conlon:2024uob} (see also \cite{Revello:2024gwa, Brunelli:2025ems, Ghoshal:2025tlk}; for earlier work on strings with time-varying tensions, see \cite{Yamaguchi:2005gp, Ichikawa:2006rw, Cheng:2008ma, Sadeghi:2009wx, Wang:2012naa, Emond:2021vts}) that, when the tension decreases, string loops grow in physical coordinates and, during an epoch of volume kination, loops of fundamental strings grow even in comoving coordinates. 
\\
\\
In this paper we continue to investigate the dynamics of a population of fundamental string loops in the background of an evolving volume modulus. 
In contrast with previous work \cite{Conlon:2024uob,Revello:2024gwa, Brunelli:2025ems, Ghoshal:2025tlk}, we will study the back-reaction due to the population of cosmic string loops on the modulus evolution and the background cosmology. In particular, we perform a dynamical systems analysis and identify a new attractor solution in the form of a string loop tracker. With a different physical scenario in mind, a mathematical analysis of the dynamical systems discussed in the next sections can also be found in \cite{Amendola:1999qq,Amendola:1999er} together with the corresponding stability analysis. In this paper we identify an additional line of fixed points in Section \ref{radalso} for a particular relation between parameters.

\subsection*{Review: Volume Kination and Time-dependent String Tensions}

In \cite{Conlon:2024uob} we studied the equations of motion for a Nambu-Goto string with time-dependent tension. Restricting to the simple case of circular loops, we showed that the physical length of the loop evolves as
\begin{equation}
\label{lgrowth}
    \ell(t) = \ell_i \sqrt{\frac{\mu_i}{\mu (t)}},
\end{equation}
with $\ell_i$ and $\mu_i$ the initial length and tension (we can view
$\ell$ as the length of the loop at the maximal point of each oscillation). From Eq. (\ref{lgrowth}), it follows that a decreasing tension leads to a growth in the physical size of the loop.\footnote{This is similar to the intuition of a harmonic oscillator with a decreasing string constant.} In the case that
\be
2 H + \frac{\dot{\mu}}{\mu} <0, \label{Hmudot}
\ee
where $H \equiv \frac{\dot{a}}{a}$ is the Hubble parameter, with $a(t)$ the scale factor, and $\dot{a} \equiv \frac{da}{dt}$ etc.., the loops grow even in comoving coordinates, implying that an isolated population of small loops will eventually find each other and percolate provided Eq. \eqref{Hmudot} holds for long enough. 
\\
\\
Each loop has an associated energy set by the product of its length and its tension,
\be
E_{\rm loop} = \mu(t) \cdot \ell(t),
\ee
with an overall energy density for a population of isolated sub-horizon loops set by
\be
\label{rhoisolatedloops}
\rho_{\rm loops}(a) = n \cdot \mu \cdot \ell = n_i \left( \frac{a_i}{a} \right)^3 \mu_i \ell_i \sqrt{ \frac{\mu(t)}{\mu_i}}.
\ee
Although loops slowly evaporate through emission of gravitational waves, on timescales much shorter than their decay time, the energy density in fixed-tension loops redshifts as matter $\rho \propto \left( \frac{a_i}{a} \right)^3$.\footnote{We assume that the rate at which loops emit heavy particles associated to e.g. the string scale is negligible. The spectrum of these particles and their coupling to the strings depends on the details of the high energy theory.} With a decreasing tension the energy density effectively redshifts faster, as $\rho \propto \left( \frac{a_i}{a} \right)^3 \sqrt{ \frac{\mu(t)}{\mu_i}}$. We analyse the effects of gravitational wave emission for the case of fundamental strings in LVS in more detail in Section \ref{Sec2.3}.
\\
\\
Detailed summaries of both volume kination and volume trackers can be found in \cite{Apers:2024ffe}. During kination, a canonically normalised scalar field $\Phi$ evolves as
\be
\label{kinevolution}
\Phi(t) = \Phi_i + M_P \sqrt{\frac{2}{3}} \ln \left( \frac{t}{t_i} \right).
\ee
The relationship between the canonically normalised volume modulus $\Phi$ and the Calabi-Yau volume $\mathcal{V}$ is
\be
\Phi = \Phi_i + M_P \sqrt{\frac{2}{3}} \ln \left( \mathcal{V} / \mathcal{V}_i \right),
\ee
($\mathcal{V}$ is a dimensionless volume, measured in the Einstein frame and expressed in units of the string length $l_s$, via $l_s^6 \equiv \left( 2 \pi \sqrt{\alpha'} \right)^6$). It follows that during kination, the volume evolves as
\be
\mathcal{V}(t) \propto t \sim a^3.
\ee
As the tension of fundamental strings behaves as 
\be
\mu_s \sim m_s^2 \sim \frac{M_P^2}{\mathcal{V}},
\ee
during a kination epoch
\be
\label{rholoopskination}
\rho_{\rm loops}(a) \sim a^{-9/2}.
\ee
From Eq. \eqref{rholoopskination}, it is clear that the energy density in loops falls less quickly than the background (which has $\rho_{\rm kination} \sim a^{-6}$) and so the loops will, given time, catch up with the background.
\\
\\ 
This motivates a dynamical systems analysis of scalar fields on exponential potentials (\cite{Copeland:1997et}; and e.g. \cite{Brunelli:2025ems} for a very recent treatment), whilst incorporating a population of (fundamental) string loops in the background. Note the loops cannot be treated as a barotropic fluid as they do not have a fixed equation of state because their redshift with the scale factor depends on the evolution of the volume modulus. 

\section{Dynamical Systems Analysis Without Radiation}\label{Sec3}

In this section, we conduct a dynamical systems analysis to determine the evolution and fixed points of an FLRW universe filled with a scalar field $\Phi$ on an exponential potential $V(\Phi)$, in the presence of an energy density $\rho(a,\Phi)$ in the form of cosmic superstring loops whose tension depends on the modulus $\Phi$. In subsection \ref{norad} we consider this system without any additional radiation, whereas in section \ref{radalso} we further include a radiation background \textcolor{blue}{(see also \cite{Amendola:1999qq,Amendola:1999er})}. 
\\
\\
In the absence of strings loops, such systems are well studied \cite{Copeland:1997et}. The system can end up in a kination epoch, in which all the energy is in the kinetic energy of the scalar field, or a tracker solution, in the cases either where the potential $V(\Phi)$ is insufficiently steep (such that a sizable fraction of the energy density is in the form of potential energy) or where a background of radiation comes to have comparable energy density to the kinating field, slowing this down by Hubble friction. 
\\
\\
We will see that the introduction of a population of cosmic string loops results in interesting new tracker solutions in which an order one fraction of the energy is in the form of string loops. For example, in the case of a background of fundamental strings in the LVS potential, the attractor solution involves three-quarters of the energy density being in fundamental strings. 
\\
\\
In our present work, we remain agnostic about the initial origin of such a population of cosmic string loops and instead focus on studying the consequences that such a pre-existing population would have on the evolution of the background. Furthermore, as a first approximation, we do not consider effects such as interactions between separate loops or the production of additional loops through interactions among an existing network of long strings (these effects could be studied by introducing additional terms in Eq. \eqref{EqRho} for the evolution of the energy density in loops).
\\
\\
As further simplifications we assume that the population of loops can be characterised by a single typical loop length and that the  spatial distribution of the population is homogeneous enough to be described by a homogeneous energy density such that we analyse only the modulus' zero momentum mode.
\subsection{Friedmann equations}
\label{norad}
Sub-horizon cosmic string loops oscillate rapidly relative to the Hubble time (recall the oscillation frequency is set by their physical size, which is typically much less than the horizon). This allows for time-averaging of the velocity over oscillations, $\langle v^2\rangle =1/2$, where the RMS velocity is defined as $v^2 \equiv \int \dot{\vec{x}}^2 \varepsilon d\sigma /\left(\int \varepsilon d\sigma \right)$ with $\sigma$ denoting the spatial coordinate on the worldsheet and the length of the loop corresponding to $\ell =\int \varepsilon d\sigma $ . The loops are also similar to matter in that they do not have a pressure ($p=0$), although the time-dependent tension means that the `mass' of a loop evolves with time. \\
\\
We can parametrise the dependence of the string loop tension $\mu$ and string loop energy density $\rho$ on the modulus $\Phi$ as,
\begin{align}
    \mu (\Phi) &= M_P^2 \; e^{-\sqrt{6} \beta \Phi/M_P},
    \\
    \rho_{\rm loops}(a,\Phi) &= n\cdot \mu \cdot \ell = \rho_{\rm loops,i} \cdot \left(\frac{a_i}{a}\right)^{-3} \cdot e^{-\sqrt{3/2} \beta \;  (\Phi - \Phi_i)/M_P}.
\end{align}
where $\beta$ is a numerical parameter fixed by the underlying theory and we have used the fact that the length changes due to the evolution of the tension as in Eq. \eqref{lgrowth}, with $\mu$ and hence $\ell$ now functions of $\Phi$ (the evolution of individual loops of fundamental strings in this background was studied in \cite{Conlon:2024uob}). For the case of fundamental strings, $\beta = 1/2$, as their tension is set by the string scale, $\mu \sim m_s^2$ with $m_s \sim M_P / \sqrt{\mathcal{V}}$. 
\\
\\
As mentioned at the beginning of this section, and also anticipated in Eq. \eqref{rhoisolatedloops}, we assume that loops are neither produced nor destroyed, so that any dilution in their number density comes solely from the expansion of the universe. We consider the case where the scalar field is the canonically normalised volume modulus with an exponential potential,
\be \label{eq:Vphi}
V(\Phi) = M_P^4 \; e^{-\sqrt{3/2}\;\lambda \Phi/M_P}.
\ee
$\lambda = 3$ corresponds to the LVS case where the potential falls off as $\mathcal{V}^{-3}$. 
\\
\\
The Friedmann and continuity equations for this energy content are
\begin{empheq}[left={\empheqlbrace}]{alignat=2}
    3M_P^2 H^2 &= \frac{\dot{\Phi}^2}{2} + V(\Phi) + \rho_{\rm loops}(a,\Phi), \label{Friedman1}
    \\
    -2M_P^2 \dot{H} &= \dot{\Phi}^2 + \rho_{\rm loops}(a,\Phi), \label{Friedman2}
    \\
    0 &= \ddot{\Phi} + 3H\dot{\Phi} + \frac{\partial V(\Phi)}{\partial \Phi} + \frac{\partial \rho_{\rm loops}} {\partial \Phi}, \label{EqPhi}
    \\
     0 &= \dot{\rho}_{\rm loops} + 3H\rho_{\rm loops} - \frac{\partial \rho_{\rm loops}} {\partial \Phi} \dot{\Phi}, \label{EqRho}
\end{empheq}
where $a(t)$ and $\Phi(t)$ are to be solved for. 

\subsection{Finding the fixed points}
We search for tracker solutions, i.e. fixed points of the system of equations Eq. \eqref{Friedman1} to \eqref{EqRho}. To that end, following the treatment in \cite{Copeland:1997et}, we define the variables:
\begin{align}
    x^2 &\equiv \dot{\Phi}^2/6M_P^2H^2, \label{defx}
    \\
    y^2 &\equiv V /3M_P^2H^2,
    \\
    z^2 &\equiv \rho_{\rm loops}/3M_P^2H^2.
\end{align}
These represent the fraction of total energy in the different forms, and so  
the Friedmann equation, Eq. \eqref{Friedman1}, can be re-expressed as
\begin{equation}
    x^2+y^2+z^2 =1.\label{xyz:constraint}
\end{equation}
A fixed point is defined as a solution for which  $x$, $y$ and $z$ are constant with time. To find these, it is sufficient to solve for $x'=y'=0$, from which it follows from Eq.~(\ref{xyz:constraint}) that $z'=0$, where (following \cite{Copeland:1997et}) the prime denotes a derivative with respect to the number of e-folds $N=\log a$ (related to the time derivative by $d/dt=H \;d/dN$).
\\
\\
Using Eq. \eqref{Friedman2} and Eq. \eqref{EqPhi} together with the relations for the potential and loop energy density $(\partial_\Phi V)/V= -\sqrt{3/2}\;\lambda/M_P$ and $(\partial_\Phi \rho_{\rm loops})/\rho_{\rm loops} = -\sqrt{3/2}\;\beta /M_P$, we obtain the following equations for the evolution of the different components:
\begin{align}
    x' &= \frac{3}{2} \left[x(x^2 -y^2 -1) + (\beta - \beta x^2 + (\lambda - \beta) y^2)\right],
    \\
    y'&=\frac{3}{2} y \left[x^2-y^2-\lambda x+1\right],
\end{align}
with $z^2 = 1- x^2-y^2$. The fixed points for this system are summarised in Table~\ref{tab:trackers} in terms of $\Omega_{\dot{\Phi}}\equiv x^2,\; \Omega_V \equiv y^2$ and $ \Omega_{\rm loops} \equiv z^2$. The range of parameters $\beta$ and $\lambda$  for which these fixed points are physically meaningful is determined by the condition $0 \leq \Omega_i \leq 1$ for all three components $i=\dot{\Phi}, \rho_{\rm loops}, V$. On physical grounds we also restrict to $\lambda, \beta \geq 0$ -- these correspond to a decreasing scalar potential and a decreasing tension for large field values. This behaviour is  expected as we approach the boundary of moduli space where $m_s \sim g_s M_P/\sqrt{\mathcal{V}} \rightarrow 0$ in the 4d Einstein frame, as all physical scales such as potentials and tensions must vanish in this limit. 
\\
\begin{table}[ht!]
    \centering
    \renewcommand{\arraystretch}{1.5}
    \begin{tabular}{|l|c|c|c|>{\raggedright\arraybackslash}p{0.3\linewidth}|}
        \hline
        \textbf{Fixed points} & \textbf{$\Omega_V$} & \textbf{$\Omega_{\dot{\Phi}}$} & \textbf{$\Omega_{\rm loops}$}& \textbf{Existence conditions} \\
        \hline
        \textbf{A. Kination} & $0$ & $1$ & $0$ & $ \forall \lambda, \beta$ \\ 
        
        \textbf{B. String Loop tracker} & $0$ & $\beta^2$ & $1 - \beta^2$ &  $\forall \lambda$, $\beta \leq 1$ \\ 
      
        \textbf{C. Scalar field domination}& $1 - \frac{\lambda^2}{4}$ & $\frac{\lambda^2}{4}$ & $0$ & $\forall \beta$ , $\lambda \leq 2$ \\ 
    
        \textbf{D. Mixed tracker I} & $\frac{\beta^2 - \lambda \beta +1}{(\lambda - \beta)^2}$ & $\frac{1}{(\lambda - \beta)^2}$ & $\frac{\lambda^2 - \lambda \beta -2}{(\lambda - \beta)^2}$ &  $\beta \leq 1$,  $\frac{\beta}{2} + \sqrt{\frac{\beta^2}{4} +2} \leq \lambda \leq \beta + \frac{1}{\beta}$\\ 
        \hline
    \end{tabular}
    \caption{Fixed points for the evolution of the system with a modulus field $\Phi$ with an exponential potential energy $V(\Phi)$ and a population of string loops (of energy density $\rho_{\rm loops}$) coupled to the modulus $\Phi$. The parameters $\beta$ and $\lambda$ are defined as $\lambda\equiv-\sqrt{2/3} M_P \, (\partial_\Phi V)/V$ and $\beta \equiv -\sqrt{2/3} M_P (\partial_\Phi \rho_{\rm loops})/\rho_{\rm loops}$ .}
    \label{tab:trackers}
\end{table}
\\
We now analyse the properties of these fixed points, among which there are two novel possibilities, the string loop tracker and the mixed tracker. In both, an $\mathcal{O}(1)$ fraction of the energy density of the universe lies in the form of fundamental strings. 
For the particular case of fundamental superstrings in specific well-motivated 
potentials (both LVS and a canonical potential scaling as $m_{string}^4$), we also give a more detailed discussion of stability.

\subsubsection*{A. Kination}
The first fixed point is the kination one, present for all $\lambda$ and $\beta$, in which all the energy lies in the kinetic energy of the scalar field. As well known, (e.g. see \cite{Copeland:1997et}), in the case of $\lambda \leq 2$, the potential $V(\phi)$ is sufficiently flat that the kination fixed point is unstable to perturbations towards the scalar field domination (C) fixed point, which has energy in both the potential and kinetic parts of the scalar field. 
\\
\\
Moreover, for all values of $\lambda$ the kination fixed point is also unstable if there is any initial energy density in loops. In such a case, the system is driven to the string loop or mixed trackers (B and D respectively) presented in Table~\ref{tab:trackers}. 

\subsubsection*{B. String loop tracker}
The second fixed point is novel and interesting. Energy is distributed between strings and the kinetic energy of the scalar field. This fixed point exists for $\beta \leq 1$. Otherwise, if $\beta >1$, the energy density in loops decreases too fast and so cannot remain at a constant fraction of the total energy density of the universe.
\\
\\
This fixed point exists for all values of $\lambda$. If the potential is not steep enough the string loop tracker can be unstable, with the system evolving towards the mixed tracker instead if there is some initial energy density in the form of potential energy. However, we will see that fixed point B is the global attractor in the case of fundamental strings in LVS.
\subsubsection*{C. Scalar field domination}
This is a standard fixed point that exists in the absence of string loops for potentials with $\lambda \leq 2$, with the total energy distributed between the kinetic and potential energy of the scalar field. 

\subsubsection*{D. Mixed tracker I}
We call the final fixed point the \textit{mixed} tracker because the energy is distributed between the potential energy of the scalar field, the kinetic energy of the scalar field and the loops. 
\\
\\
Similarly to the string loop tracker, it exists if $\beta \leq 1$ as otherwise the energy density in loops decreases too fast for a fixed point to exist. Additionally, for this fixed point to exist the steepness of the potential $\lambda$ must take values in the range $\frac{\beta}{2} + \sqrt{\frac{\beta^2}{4} +2} \leq \lambda \leq \beta + \frac{1}{\beta}$. If the lower bound of this inequality is violated the potential redshifts slowly enough to come to dominate over the loops. Conversely, if the upper bound is violated the potential energy decreases too fast for such a fixed point to exist. In other words, if $\lambda$ lies outside this range, a system starting from the would-be mixed tracker is unstable to either kination or the string loop tracker.
\\
\\
In the case of constant tension string loops ($\beta =0$) the mixed tracker simply corresponds to a matter tracker (see e.g. \cite{Apers:2024ffe} for a recent discussion). 

\subsection{Fundamental Strings in Specific Potentials}\label{Sec2.3}

The above analysis was for generic $\beta$ and $\lambda$. Here we look at the case of fundamental strings ($\beta = 1/2$) for some specific, motivated, examples of exponential potentials.

\subsubsection*{A canonical potential}
\begin{figure}[ht!]
    \centering
    \includegraphics[scale=0.35]{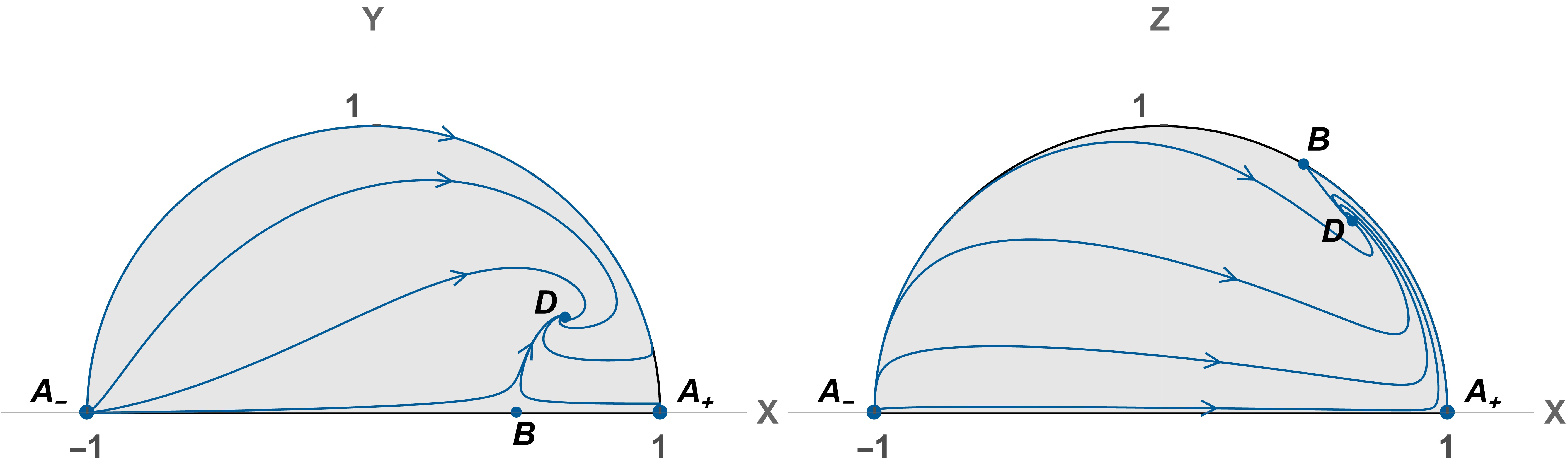}
    \caption{Evolution in phase space towards the mixed tracker labeled as point D, for the case of fundamental strings in a potential of the form $V\sim m_s^4$ , i.e. $\beta=1/2$ and $\lambda=2$. Points $A_\pm$ correspond to kination and point B to the string loop tracker}
    \label{fig:MixedTracker}
\end{figure}

The simplest type of a string theory potential driving fields to the boundary of moduli space is of the form $V(\Phi) \sim m_s^4 \sim \mathcal{V}^{-2}$, set by the natural scale of the theory.\footnote{We note that as written, this has the problem of lacking the construction of a minimum or vacua in the asymptotic region of 
moduli space (whereas the scaling $V(\Phi) \sim \mathcal{V}^{-3}$ allows for the existence of the LVS minimum at large volumes). However interesting the early cosmology is, it needs a final minimum for the modulus to end up in.} This potential corresponds to $\lambda=2$.
\\
\\
In this case, the fixed points of the system are kination and the string loop and mixed trackers. The first two are unstable, since on these the potential is not sufficiently steep to redshift away with respect to the background. In fact, the stable fixed point is the mixed tracker where the fractions in energy densities are
\begin{align}
    \Omega_V = 1/9 \hspace{2 em} \Omega_{\dot{\Phi}} = \Omega_{\rm loops}= 4/9 
\end{align}
and the string loops behave as radiation, such that $a(t) \sim t ^{1/2}$ and $G\mu(t) \sim t^{-1}$. Loops of fundamental strings will continue to grow in physical size during this tracker, but not in co-moving coordinates and hence do not percolate. Fig. \ref{fig:MixedTracker} shows the attractor behavior of this fixed point in phase space. 
\subsection*{The Large Volume Scenario (LVS)}
\begin{figure}[ht!]
    \centering
    \includegraphics[scale=0.35]{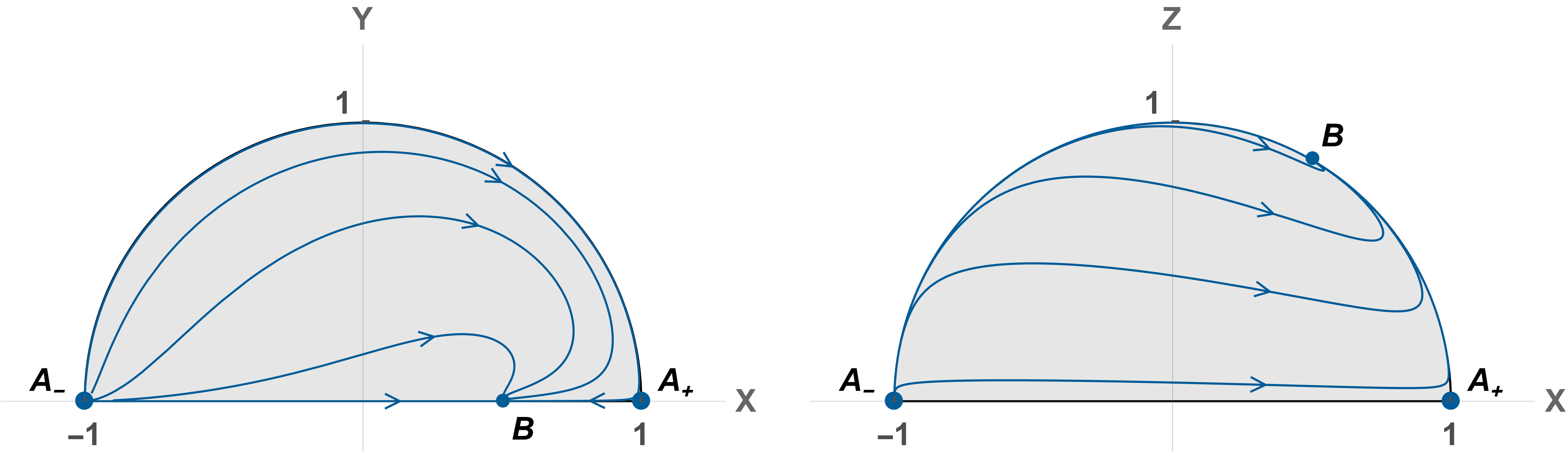}
    \caption{Evolution in phase space towards the attractor corresponding to the string loop tracker, labeled as point B for the case of fundamental strings in LVS ($\beta=1/2$ and $\lambda=3$). Points $A_\pm$ correspond to kination.}
    \label{fig:AttractorLoops}
\end{figure}

\begin{figure}[ht!]
  \centering
    \includegraphics[scale=0.5]{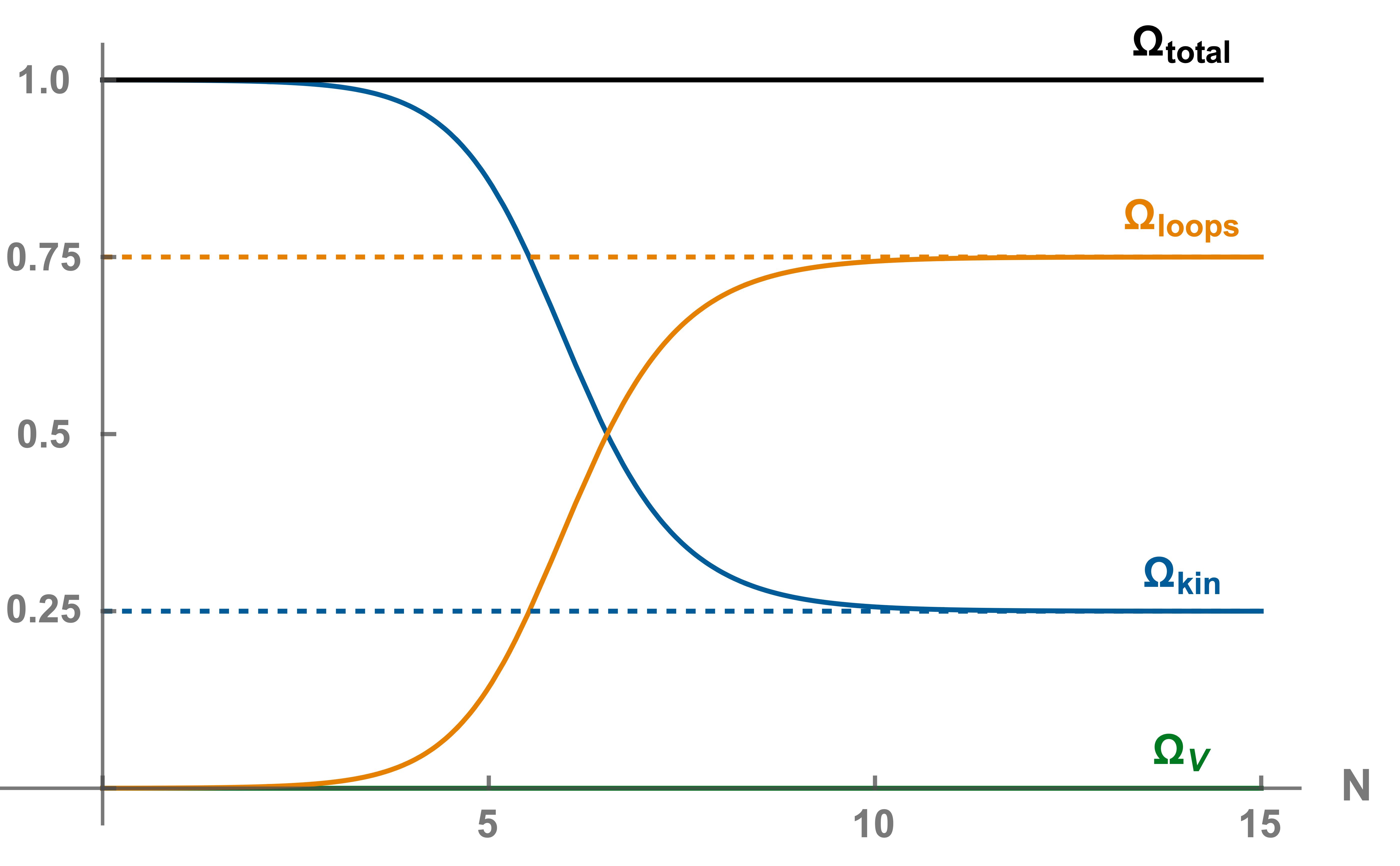}
    \caption{Evolution of a initially kinating background, with a small initial $\Omega_{\rm loops}^i=10^{-2}$, towards the string loop tracker for $\lambda =3$ and $\beta =1/2$,
    plotted in terms of the number of e-foldings $N$.}
    \label{fig:Newtracker}
\end{figure}

A particularly well-motivated case is of fundamental strings with the LVS potential.\footnote{Here we restrict ourselves to the runaway exponential part of the LVS potential that holds at small volume; the full potential has extra structure creating the minimum at large volume.}
As the LVS minimum is located at an exponentially large volume, it is at a significantly trans-Planckian distance in field space from the centre of moduli space. The large field excursions necessary for kination/tracker scenarios are therefore well motivated (arguably required) in LVS in a way they are not in other scenarios of moduli stabilisation.
\\
\\
The LVS potential corresponds to $\lambda = 3$ (this is equivalent to $V \sim \mathcal{V}^{-3}$), and fundamental strings have $\beta=1/2$ (we do not consider any subtleties such as tying the strings to warped regions of the Calabi-Yau).
\\
\\
In this case, since the potential is steep, i.e. $\lambda > 2$, the system only has two potential fixed points: either a kination epoch or a string loop tracker. The former is unstable to perturbations from the energy density of loops; in particular the moment there is any non-zero energy density in loops, this will grow with respect to the kinating background ($a^{-9/2}$ vs. $a^{-6}$) and drive the system to the string loop tracker on which $3/4$ of the energy of the universe is in string loops with the remaining $1/4$ in the kinetic energy of the modulus:
\begin{align}
    \Omega_V = 0 \hspace{2 em} \Omega_{\dot{\Phi}} = 1/4 \hspace{2 em} \Omega_{\rm loops}=3/4~.
\end{align}
Fig. \ref{fig:AttractorLoops} shows the evolution of the system in phase space as it ends in the string loop tracker (point B) which acts as an attractor. Fig. \ref{fig:Newtracker} shows a sample evolution of the system for some initial conditions from the kination fixed point to the string loop tracker. 
\\
\\
The explicit time-dependence of the scale factor and the tension of the cosmic string loops on the string loop tracker can be calculated from Eq.~\eqref{Friedman2} together with the values in Table \ref{tab:trackers}. We find that $H=\frac{8}{15t}$, which sets the expansion rate, and using Eq. \eqref{defx} evaluated at the string loop tracker fixed point given in Table \ref{tab:trackers}, we can then determine $\Phi(t)$. As a result, we find that during the string loop tracker
\begin{align}
    a(t) &\sim t ^{8/15}
    \\
    G\mu(t) &\sim t^{-4/5}
\end{align}
Given the time-dependence of the tension, the loops continue to grow in physical size during this epoch, but, unlike in kination, they shrink in comoving coordinates (this is the case as long as $\beta^2 <1/3$).
\\
\\
Although we study the dynamics in the presence of radiation in more detail in the next Section, let us note here that as $\rho_{\rm radiation} \sim a^{-4}$, during this epoch $\rho_{\rm radiation} \sim t^{-32/15}$ and so radiation redshifts faster than the background; the string loop tracker is therefore stable against a small component of radiation.\footnote{Note that matter redshifting as $\rho \sim a^{-3}$ would catch up with the tracker; this is not satisfied for particle-like excitations, as in string compactifications the masses of particles all depend on the evolving modulus. It would, however, be satisfied by any primordial black holes present.}
\\
\\
Let us consider the stability of the loops against gravitational radiation. Loops emit gravitational radiation with a power $P_{GW} = \Gamma G\mu^2$, where $\Gamma$ is a numerical factor that depends on the precise loop configuration. Nambu-Goto simulations suggest $\Gamma \sim  50 - 75$. Therefore, the order-of-magnitude lifetime of a loop of length $\ell$ and mass $\mu \ell$ is (using $\mu \sim m_s^2$),
\begin{align}
    \tau_{GW} \sim \frac{\ell}{\Gamma G \mu}= \frac{8\pi}{\Gamma} \cdot\frac{M_P^2}{m_s^2} \cdot \ell
\end{align}
Since the growth of the loop is due to the change in tension on a Hubble time, $\tau_H$, if $\tau_{GW} \gg \tau_H$, the effects of gravitational emission on its evolution will be negligible. During the string loop tracker, the string scale evolves as $m_s/M_P \sim t^{-2/5}$. Using the result that the length of the loops evolves due to the change in tension as $\ell \sim 1/\sqrt{\mu} \sim t^{2/5}$, we can see that the ratio between the gravitational emission lifetime and the Hubble timescale will increase with time,
\begin{align}
    \frac{\tau_{GW}}{\tau_H} \sim \frac{8\pi}{\Gamma}\cdot  \frac{M_P^2}{m_{s,i}^2}\cdot  \ell_i  H_i  \cdot \left(\frac{t}{t_i}\right)^{1/5} ,
\end{align}
and so the loops become progressively \emph{more} stable with time during the evolution along the string loop tracker (here $i$ denotes quantities at the start of the tracker).
There will therefore exist a critical size for the loops $\ell_{\rm crit}$ above which they do not evaporate at the start of the tracker, 
\begin{align} \label{crit_size}
    \ell_i > \ell_{crit}\sim \left(\frac{m_{s,i}}{M_P}\right)^2 H_i^{-1}.
\end{align}
If this is satisfied, so that we initially have $\tau_{GW} > \tau_H$, then during the string loop tracker the  subsequent back-reaction of the gravitational emission on the growth of the loops becomes progressively less important and the loops remain stable against gravitational wave emission. 

\subsubsection*{Late-Time Evolution} 

The string loop tracker is an interesting solution because as the universe enters a moduli domination epoch, where the modulus settles into the minimum and transfers its energy to SM degrees of freedom, this population of loops will survive for some time, several Hubble times in the case of a small tension.
\\
\\
There does however remains the question of how the modulus could transition from this tracker solution to oscillating around the minimum of its potential. In particular, this is problematic since on the tracker the potential energy is negligible, so there is a danger that the scalar field will not slow down enough and will instead overshoot \cite{Conlon:2022pnx}. However, we also know that in reality there is probably some energy density in radiation at the start of the evolution. As we will see next, this can alter the dynamics.

\section{Analysis With Background Radiation}
\label{radalso}
The case of strictly zero radiation is unphysical -- any physical system will radiate some quantity of energy, and the loops themselves radiate energy in the form of gravitational waves. Therefore, we now extend the dynamical systems analysis to include a radiation fluid, with $p_{\rm rad} = \frac{1}{3} \rho_{\rm rad}$.\footnote{We assume that the energy density in radiation is dominated by a pre-existing component, rather than gravitational waves produced by the loops during the era of interest (which would enter the evolution equations in a more complicated way). At a fixed point with $\rho_{\rm loops}\neq 0$ such gravitational waves would be continually sourced, but are unimportant provided the fixed point is stable to radiation perturbations.} The corresponding Friedmann and continuity equations are
\begin{empheq}[left={\empheqlbrace}]{alignat=2}
    3M_P^2 H^2 &= \frac{\dot{\Phi}^2}{2} + V(\Phi) + \rho_{\rm loops}(a,\Phi) +\; \rho_{\rm rad}(a) ~,
    \\
    -2M_P^2 \dot{H} &= \dot{\Phi}^2 + \rho_{\rm loops}(a,\Phi) +\; \frac{4}{3}\rho_{\rm rad}(a) ~,
    \\
    0 &= \ddot{\Phi} + 3H\dot{\Phi} + \frac{\partial V(\Phi)}{\partial \Phi}+ \frac{\partial \rho_{\rm loops}} {\partial \Phi} ~,\label{EqRhophi} 
    \\
    0 &= \dot{\rho}_\ell + 3H\rho_\ell - \frac{\partial \rho_{\rm loops}} {\partial \Phi} \dot{\Phi} ~, \label{EqRholoops} 
    \\
    0 &\;= \dot{\rho}_{\rm rad} + 4H\rho_{\rm rad}~. \label{EqRhorad}
\end{empheq}
\subsection{Finding the fixed points}
To search for the fixed points of the tracker solutions, we define a similar change of variables to before with an additional variable $w$ corresponding to the energy density in radiation:
\begin{align}
    x^2 &\equiv \dot{\Phi}^2/6M_P^2H^2,
    \\
    y^2 &\equiv V /3M_P^2H^2,
    \\
    z^2 &\equiv \rho_{\rm loops} /3M_P^2H^2,
    \\
     w^2 &\equiv \rho_{\rm rad} /3M_P^2 H^2,
\end{align}
such that the constraint given by the Friedman equation is now $x^2+y^2+z^2+w^2=1$. In terms of these variables, and still considering theories in which $\Phi$ has an exponential potential given by Eq. \eqref{eq:Vphi}, the equations of motion become (with derivatives with respect to $N = \ln a$) 
\begin{align}
    x' &= \frac{3}{2} \left[x(x^2 -y^2 \; +\;\frac{w^2}{3}-1) + \beta(1 - x^2 -\;w^2)+ (\lambda - \beta) y^2)\right],
    \\
    y'&=\frac{3}{2} y \left[x^2-y^2\; +\;\frac{w^2}{3}-\lambda x+1\right],
    \\
     w'&= \frac{3}{2} w \left[x^2-y^2+\frac{1}{3}(w^2-1)\right],
\end{align}
with $z^2 = 1- x^2-y^2-w^2$. These equations describe the system with the modulus and a background of both string loops and radiation.
\\
\\
The fixed points for the absence of radiation previously found in Table \ref{tab:trackers} will remain fixed points of this system when $w=0$. However, their stability now has to be reevaluated given the presence of additional fixed points with some energy fraction in radiation. These additional fixed points are summarised in Table~\ref{tab:trackersRad}. 
\begin{table}[ht!]
    \centering
    \renewcommand{\arraystretch}{1.5}
    \begin{tabular}{|l|c|c|c|c|>{\raggedright\arraybackslash}p{0.15\linewidth}|}
        \hline
        \textbf{Fixed points} & \textbf{$\Omega_V$} & \textbf{$\Omega_{\dot{\Phi}}$} & \textbf{$\Omega_{\rm{loops}}$} & $\Omega_{\text{rad}}$ & \textbf{Existence conditions} \\
        \hline
        \textbf{E. Radiation Tracker}& $\frac{8}{9\lambda^2}$ & $\frac{16}{9\lambda^2}$ &  $0$ & $1-\frac{8}{3\lambda^2}$ & $\forall \beta \neq
        \frac{\lambda}{4} $, $2\sqrt{\frac{2}{3}} \leq \lambda$\\ 
        \textbf{F. Loop-Radiation Tracker}& $y_{\rm fp}^2$ & $\frac{16}{9\lambda^2}$ &$\frac{32}{9\lambda^2} -4y_{\rm fp}^2$&   $1-\frac{16}{3\lambda^2} +3y_{\rm fp}^2$& $\beta = \frac{\lambda}{4}$ , $2 \sqrt{\frac{2}{3}}  \leq \lambda$ \\
        \textbf{G. Mixed Tracker II} & $0$ & $\frac{1}{9\beta^2}$ & $\frac{2}{9\beta^2}$ &   $1-\frac{1}{3\beta^2}$ & $\forall \lambda$, $\frac{1}{\sqrt{3}} \leq \beta$\\ 
        \textbf{H. Radiation Domination}& $0$& $0$& $0$& $1$&$\forall \lambda, \beta$\\ 
        \hline
    \end{tabular}
    \caption{Fixed points in the evolution of the system with a scalar field $\Phi$ on an exponential potential $V(\Phi)$, radiation $\rho_{\rm rad}$ and a population of string loops (energy density $\rho_{\rm loops}$) coupled to the modulus $\Phi$. The parameters $\beta$ and $\lambda$ are defined as $\lambda\equiv-\sqrt{2/3} M_P (\partial_\Phi V)/V$ and $\beta \equiv -\sqrt{2/3} M_P (\partial_\Phi \rho_{\rm loops})/\rho_{\rm loops}$.}
    \label{tab:trackersRad}
\end{table}
\\
\\
The fixed point E, on which there is no energy density in string loops, is the well-known tracking solution of Ref. \cite{Copeland:1997et}. In contrast, the fixed points F and G correspond to tracker solutions in which the energy density of the universe is in the form of both radiation and string loops. The mathematical analysis of this system was also done in \cite{Amendola:1999er}, with the addition here of the line of fixed points labeled by the letter F which we discuss in more detail below. Note that it appears due to a degeneracy in phase space for the particular relation between parameters $\beta=\lambda/4$.
\subsubsection*{E. Radiation Tracker }

In the absence of the string loops fixed point E, the well-known tracker solution of \cite{Copeland:1997et}, is an attractor solution. However, this is not automatic in the presence of strings loops. The stability of the fixed point in this case can be determined from the evolution of the individual energy densities, 
\begin{align}
    \rho_{\rm rad} &\sim a^{-4} ,
    \\
    \rho_{\rm loops} &\sim  a^{-3} \cdot \mathcal{V}^{-\beta} ,
    \\
    V  &\sim \mathcal{V}^{-\lambda} .
\end{align}
As on this tracker the energy density in radiation remains at a 
constant ratio to the background, we know that $a\sim t^{1/2}$. Since the 
fraction in the potential energy also remains constant relative to the background, it also follows that $\mathcal{V} \sim t^{2/\lambda} \sim a^{4/\lambda}$, which implies
\begin{align}
    \rho_{\rm loops} &\sim  t^{-3/2} \cdot t^{-2\beta/\lambda} \sim a^{-(3+4\beta/\lambda)} ~.
\end{align}
As a result, when $\beta <\lambda/4$, the string loops redshift slower than radiation and the pure-radiation tracker is unstable (we will see this holds for fundamental strings in LVS). On the other hand, if $\beta > \lambda/4$, string loops redshift faster than radiation and therefore do not destabilize the system once it gets to this attractor. Instead, string loops are diluted relative to the background by the cosmological expansion (which is driven by the equation of state corresponding to radiation). 
\\
\\
 We consider the boundary case $\beta = \lambda /4$, which we call loop-radiation tracker, next.

\subsubsection*{F. Loop-Radiation Tracker}
For the specific value $\beta =\lambda/4$, loops redshift the same way as radiation. Moreover, if the potential energy is a constant fraction of the total energy, we have $V(\Phi)\sim \mathcal{V}^{-\lambda}\sim t^{-2}$, which for $\beta=\lambda/4$ implies that the energy density in loops also remains constant with respect to the background,  $\rho_{\rm loops} \sim t^{-2}$. 
In fact for such $\beta$ and $\lambda$, there is a line of fixed points along which the relative proportion of energy in loops and radiation is a free parameter (the fraction of energy in $\Phi$'s potential also varies along this line because the energy density in loops enters into the equation of motion of $\Phi$, Eq. \eqref{EqRhophi}). These fixed points can be parameterised by $y_{\rm fp}$, the value of $y$ at the fixed point, which has to lie in the range $y_{\rm fp,\min}^2 \leq y_{\rm fp}^2 \leq 8/(9\lambda^2)$. For $2 \leq \sqrt{3/2} \;\lambda \leq 2\sqrt{2}$, $y^2_{\rm \rm fp,\min}(\lambda) = 16/(9\lambda^2)-1/3$ and for $2 \sqrt{2} \leq \sqrt{3/2}\lambda $, $y^2_{\rm fp,\min} = 0$. 
We therefore have a continuous family of tracker solutions with energy distributed between radiation, string loops, and modulus potential and kinetic energy. In the context of dynamical systems, such a line of fixed points would be called a 1-dimensional critical manifold \cite{wainwright1997dynamical,Bahamonde:2017ize}.
\\
\\
In this case, the relative fractions of energy densities in radiation, potential and string loops are set by the initial conditions: once a tracker solution is reached there is a flat direction in phase space. The parameters for this line of fixed points to exist are realised for fundamental strings in a potential like $V \sim m_s^4 \sim \mathcal{V}^{-2}$.

\subsubsection*{G. Mixed Tracker II}
There is also a fixed point in which there is no fraction of energy in the modulus' potential. Here, the equation of state corresponds to radiation, $a\sim t^{1/2}$, and the modulus evolves such that the loops behave like radiation since $\mathcal{V} \sim t^{1/(2\beta)}$ (in this case, the interplay of the loop energy density and the evolution of the modulus results in there not being a line of fixed points). For this fixed point to exist, the loops need a strong enough dependence on the modulus, $\beta \geq 1/\sqrt{3}$. 

\subsubsection*{H. Radiation Domination}
Radiation domination is also a fixed point. However, this is always unstable: as the Hubble scale decreases, the modulus will start rolling again on the exponential potential and also, while the modulus is frozen, the energy density in loops behaves like ordinary matter $a^{-3}$ and so redshift slower than the background. 

\subsection{Fundamental strings in specific potentials}

As for the case without radiation, we now study in more detail the evolution of fundamental strings (with $\beta=1/2$) in some specific and motivated forms of exponential potentials.

\subsubsection*{A canonical potential}
\begin{figure}[ht!]
    \centering
    \includegraphics[scale=0.35]{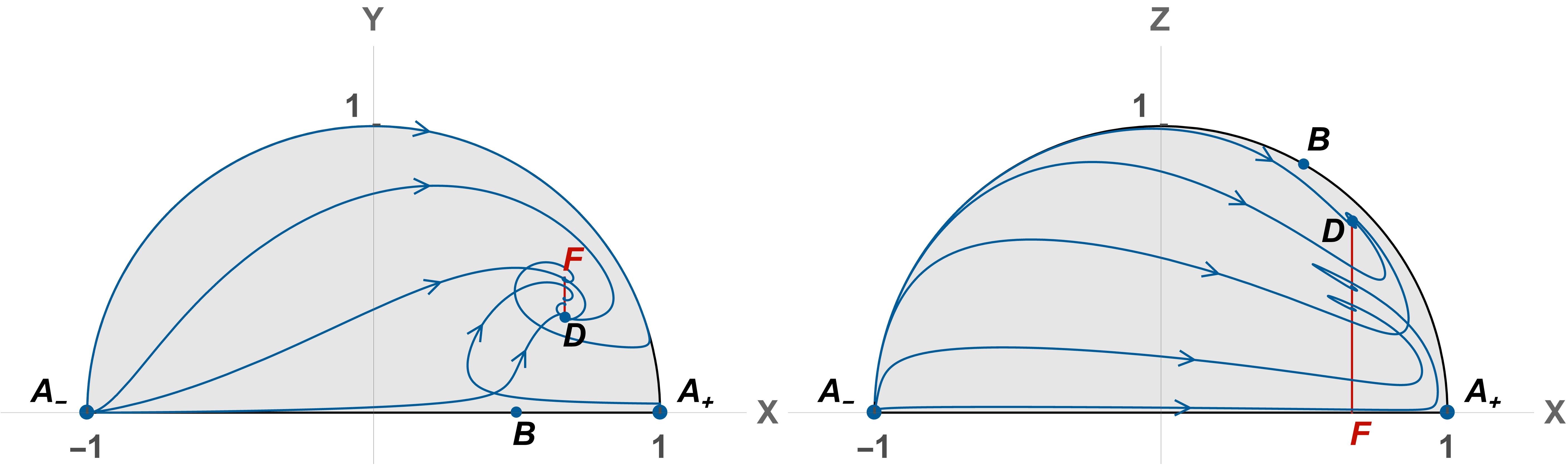}
    \caption{Evolution in phase space towards a loop-radiation tracker (red line, labeled F) for fundamental strings in a potential of the form $V\sim m_s^4$, i.e. $\beta=1/2$ and $\lambda =2$. This tracker corresponds to a line rather than a point because there is a flat direction in phase space.}
    \label{fig:AttractorRadII}
\end{figure}

The case $\lambda=2$ corresponds to a potential set by the string scale, $V(\Phi) \sim m_s^4$. Interestingly, this corresponds to the particular case $\lambda = 4 \beta$ that appears in Table \ref{tab:trackersRad} with scenario F being the fixed point. 
\\
\\
In this case, this loop-radiation tracker is the final, stable, attractor, and is characterised by all components redshifting as radiation, but with a flat direction for the relative fractions of energy density in loops and radiation. The evolution in phase space for some sample initial conditions close to kination is shown in Fig. \ref{fig:AttractorRadII}. 
\subsubsection*{The Large Volume Scenario}
\begin{figure}[ht!]
    \centering
    \includegraphics[scale=0.35]{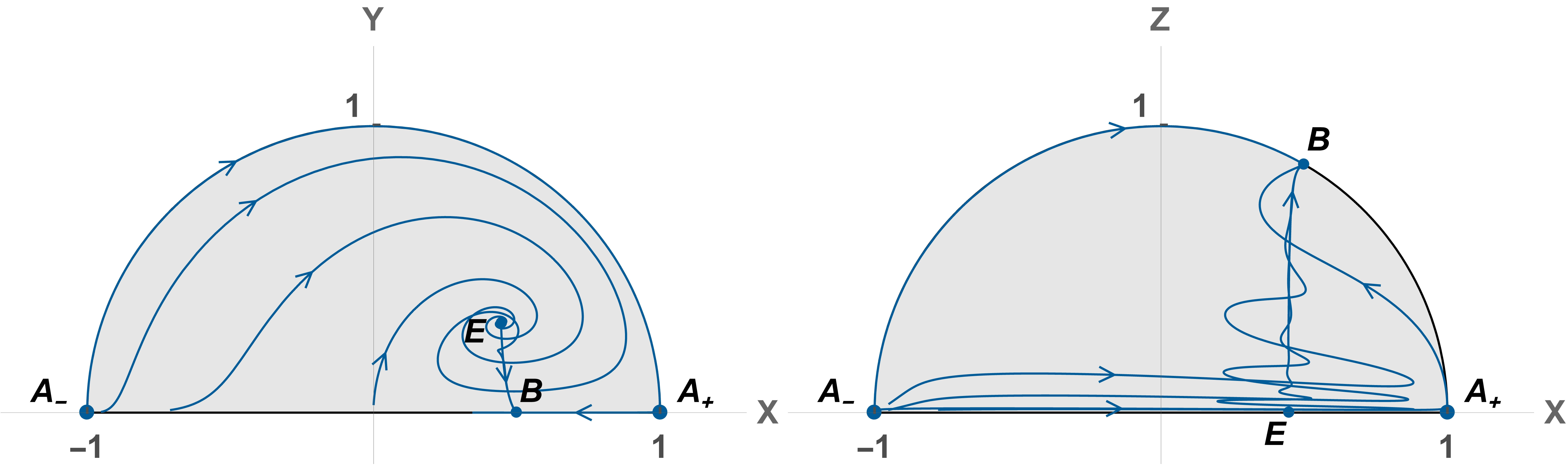}
    \caption{Evolution in phase space towards the loop tracker attractor  (B) through a transient radiation tracker (E) for fundamental strings in LVS (with $\beta=1/2$ and $\lambda =3$).}
    \label{fig:PhaseSpace_Rad}
\end{figure}

\begin{figure}[ht!]
    \centering
    \includegraphics[scale=0.5]{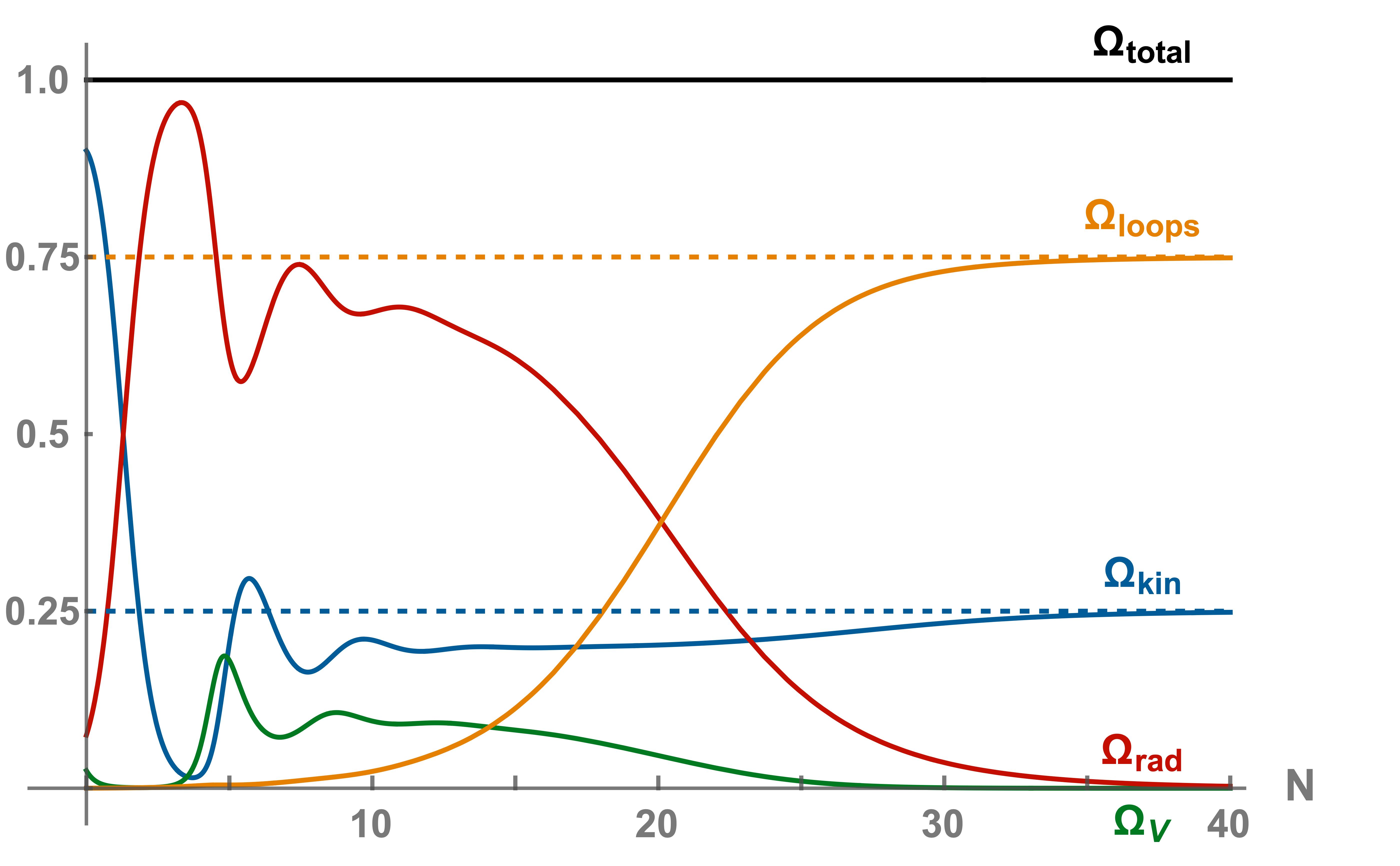}
    \caption{Evolution of the energy fractions starting from kination and going towards a string loop tracker through a transient radiation tracker, characterised by $\Omega_r\simeq 0.7$ and $\Omega_{\dot{\Phi}}=2\Omega_V \simeq 0.2$, for fundamental strings in LVS ($\beta=1/2$ and $\lambda =3$).}
    \label{fig:radtracker}
\end{figure}

The LVS potential corresponds to $\lambda = 3$ (this is equivalent to $V \sim \mathcal{V}^{-3}$). In the presence of radiation and in the absence of string loops, the familiar radiation tracker (fixed point E) would be an attractor for the system. However, in the presence of fundamental string loops with $\beta=1/2$, the fact that $\beta < \lambda/4$ implies that string loops blueshift relative to the background, making the radiation fixed point unstable. 
\\
\\
For such $\lambda$ and $\beta$, during an initial kination epoch, radiation redshifts slower than the energy density in loops ($a^{-4}$ vs $a^{-9/2}$), because the energy density in loops is affected by the fast-rolling scalar field during the kination epoch. Therefore, unless there is a significant energy density in string loops relative to radiation at the start, radiation catches up first with the kinating background. This is followed by a transient pure radiation phase (see \cite{Apers:2024ffe} for a recent analysis in the absence of string loops),
during which epoch the loops redshift as matter with $\rho_{\rm loops} \sim a^{-3}$ (as the modulus is not evolving and so the tension is temporarily fixed) and therefore their energy density grows relative to the background. 
\\
\\
The universe subsequently approaches the radiation tracker, and the scalar field starts evolving again. The energy density in loops subsequently evolves as $\rho_{\rm loops} \sim a^{-11/3}$ such that $\rho_{\rm loops}/\rho_{\rm tot} \sim t^{1/6}$,  growing slowly with respect to the energy density of the background $\rho_{\rm tot}\sim t^{-2}$. Finally, the ultimate stable attractor is the string loop tracker, in which existing radiation is slowly diluted as $\rho_{\rm rad}/\rho_{\rm tot} \sim  t^{-2/15}$. 
\\
\\
Fig. \ref{fig:PhaseSpace_Rad} shows the behavior of the radiation tracker (E) as a transient attractor, with the system eventually ending in the string loop tracker (B). In the language of dynamical systems, this transient attractor would correspond to a saddle point. It behaves as an attractor in certain directions (in the case of the radiation tracker E, that would be in the absence of loops $z=0$) but is unstable in others (such as the radiation tracker E in the presence of a non-zero density of loops $z\neq 0$). Fig. \ref{fig:radtracker} shows an example of the time evolution given some specific initial conditions, starting in a modulus kination epoch, going through a transient radiation tracker before eventually evolving towards the string loop tracker as expected. 
\subsubsection*{Late-Time evolution}
Finally, we can return to the question of late-time evolution. The maximum amount of time since kination starts until the modulus gets stabilised, due to nucleosynthesis constraints, is $t_f/t_i\sim 10^{10}$ which translates into a constraint on the number of e-folds $\Delta N\sim\mathcal{O}(1-10)$. 
\\
\\
It is therefore reasonable that in the process of going through a radiation tracker and reaching the string loop tracker (which acts as a weak final attractor) the potential energy has not become completely negligible once the modulus reaches its minimum, making moduli stabilisation possible. 
We leave a full analysis to upcoming work: from our current perspective, the key new possibility given by the existence of the string loop tracker solution is that the energy in loops can be an order one of the energy of the universe once moduli domination starts. 

\section{Conclusions}\label{Conclusions}

The most important result of this paper is the existence of a string loop tracker: a configuration in which three-quarters of the energy density of the universe lies in the form of fundamental superstring loops. In the context of the volume modulus rolling down the LVS potential, this is the stable attractor solution and is preferred over radiation tracker solutions in which radiation balances the rolling modulus.
\\
\\
In the string loop tracker, the strings are in the form of sub-horizon loops (not long strings or a network, e.g. as considered in \cite{Revello:2024gwa}). While not a thermodynamic state as such, a configuration in which the majority of the energy lies in the form of fundamental strings bears some similarity to Hagedorn phases (e.g. as recently studied in \cite{Frey:2023khe, Frey:2024jqy}, and for earlier approaches see \cite{Mitchell:1987hr,Mitchell:1987th,Atick:1988si}). We find it appealing that such a string-dominated configuration can naturally arise in a post-inflationary epoch of the universe and that it does not require a cosmological singularity or a breakdown of the notions of space and time.
\\
\\
The time-dependence of the string tension is crucial in ensuring that sub-horizon loops remain stable against their gravitational emission. In the constant tension case, sub-horizon loops behave like ordinary matter, which can also catch up with a kinating background and lead to a matter tracker. However, the constant tension sets their lifetime, and they eventually decay through gravitational emission. On the string loop tracker, the continual reduction in tension of the sub-horizon loops implies a continual reduction in the rate at which they radiate energy, resulting in their stability.
\\
\\
Our work in this paper has focused on the period during which a modulus rolls down the exponential potential. This terminates as the modulus reaches the final minimum. The modulus then oscillates about the minimum and decays, initiating the Hot Big Bang. In future work we will study the cosmological and observational consequences, including the gravitational wave signal, that arises if the string loop tracker is reached prior to the modulus attaining the minimum. We also leave for future work the study of additional effects associated to a spatial dependence in the energy density of the loop population.

\acknowledgments

We thank Fien Apers, Martin Mosny, Filippo Revello, Gonzalo Villa for discussions related to this topic. JC acknowledges support from the STFC consolidated grants ST/T000864/1 and ST/X000761/1 and is also a member of the COST Action COSMIC WISPers CA21106, supported by COST (European Cooperation in Science and Technology). EJC acknowledges support from the STFC Consolidated Grant ST/X000672/1. EH acknowledges the UK Research and Innovation Future Leader Fellowship MR/V024566/1. NSG acknowledges support from the Oxford-Berman Graduate Scholarship jointly funded by the Clarendon Fund and the Rudolf Peierls Centre for Theoretical Physics Studentship. For the purpose of Open Access, the authors have applied a CC BY public copyright licence to any Author Accepted Manuscript version arising from this submission.

\newpage
\bibliography{AlltheRefs}

\providecommand{\href}[2]{#2}\begingroup\raggedright\begin{thebibliography}{10}

\bibitem{Cicoli:2023opf}
M.~Cicoli, J.~P. Conlon, A.~Maharana, S.~Parameswaran, F.~Quevedo, and I.~Zavala, {\it {String cosmology: From the early universe to today}},  {\em Phys. Rept.} {\bf 1059} (2024) 1--155, [\href{http://arxiv.org/abs/2303.04819}{{\tt arXiv:2303.04819}}].

\bibitem{Brandenberger:2023ver}
R.~Brandenberger, {\it {Superstring cosmology \textemdash{} a complementary review}},  {\em JCAP} {\bf 11} (2023) 019, [\href{http://arxiv.org/abs/2306.12458}{{\tt arXiv:2306.12458}}].

\bibitem{Conlon:2022pnx}
J.~P. Conlon and F.~Revello, {\it {Catch-me-if-you-can: the overshoot problem and the weak/inflation hierarchy}},  {\em JHEP} {\bf 11} (2022) 155, [\href{http://arxiv.org/abs/2207.00567}{{\tt arXiv:2207.00567}}].

\bibitem{Revello:2023hro}
F.~Revello, {\it {Attractive (s)axions: cosmological trackers at the boundary of moduli space}},  {\em JHEP} {\bf 05} (2024) 037, [\href{http://arxiv.org/abs/2311.12429}{{\tt arXiv:2311.12429}}].

\bibitem{Shiu:2023fhb}
G.~Shiu, F.~Tonioni, and H.~V. Tran, {\it {Late-time attractors and cosmic acceleration}},  {\em Phys. Rev. D} {\bf 108} (2023), no.~6 063528, [\href{http://arxiv.org/abs/2306.07327}{{\tt arXiv:2306.07327}}].

\bibitem{Shiu:2023nph}
G.~Shiu, F.~Tonioni, and H.~V. Tran, {\it {Accelerating universe at the end of time}},  {\em Phys. Rev. D} {\bf 108} (2023), no.~6 063527, [\href{http://arxiv.org/abs/2303.03418}{{\tt arXiv:2303.03418}}].

\bibitem{Seo:2024qzf}
M.-S. Seo, {\it {Asymptotic behavior of saxion-axion system in stringy quintessence model}},  \href{http://arxiv.org/abs/2403.07307}{{\tt arXiv:2403.07307}}.

\bibitem{Apers:2024ffe}
F.~Apers, J.~P. Conlon, E.~J. Copeland, M.~Mosny, and F.~Revello, {\it {String theory and the first half of the universe}},  {\em JCAP} {\bf 08} (2024) 018, [\href{http://arxiv.org/abs/2401.04064}{{\tt arXiv:2401.04064}}].

\bibitem{Shiu:2024sbe}
G.~Shiu, F.~Tonioni, and H.~V. Tran, {\it {Analytic bounds on late-time axion-scalar cosmologies}},  {\em JHEP} {\bf 09} (2024) 158, [\href{http://arxiv.org/abs/2406.17030}{{\tt arXiv:2406.17030}}].

\bibitem{Andriot:2024jsh}
D.~Andriot, S.~Parameswaran, D.~Tsimpis, T.~Wrase, and I.~Zavala, {\it {Exponential quintessence: curved, steep and stringy?}},  {\em JHEP} {\bf 08} (2024) 117, [\href{http://arxiv.org/abs/2405.09323}{{\tt arXiv:2405.09323}}].

\bibitem{Conlon:2024uob}
J.~P. Conlon, E.~J. Copeland, E.~Hardy, and N.~S. Gonz\'alez, {\it {Percolating cosmic string networks from kination}},  {\em Phys. Rev. D} {\bf 110} (2024), no.~8 083537, [\href{http://arxiv.org/abs/2406.12637}{{\tt arXiv:2406.12637}}].

\bibitem{Revello:2024gwa}
F.~Revello and G.~Villa, {\it {Cosmic (super)strings with a time-varying tension}},  \href{http://arxiv.org/abs/2411.04186}{{\tt arXiv:2411.04186}}.

\bibitem{Cicoli:2024yqh}
M.~Cicoli, F.~Cunillera, A.~Padilla, and F.~G. Pedro, {\it {From inflation to quintessence: a history of the universe in string theory}},  {\em JHEP} {\bf 10} (2024) 141, [\href{http://arxiv.org/abs/2407.03405}{{\tt arXiv:2407.03405}}].

\bibitem{Casas:2024oak}
G.~F. Casas and I.~Ruiz, {\it {Cosmology of light towers and swampland constraints}},  {\em JHEP} {\bf 12} (2024) 193, [\href{http://arxiv.org/abs/2409.08317}{{\tt arXiv:2409.08317}}].

\bibitem{Apers:2024dtn}
F.~Apers, J.~P. Conlon, and M.~Mosny, {\it {A note on 4d kination and higher-dimensional uplifts}},  {\em Eur. Phys. J. C} {\bf 85} (2025), no.~3 337, [\href{http://arxiv.org/abs/2409.08049}{{\tt arXiv:2409.08049}}].

\bibitem{Brunelli:2025ems}
L.~Brunelli, M.~Cicoli, and F.~G. Pedro, {\it {Growth of Cosmic Strings beyond Kination}},  \href{http://arxiv.org/abs/2503.11293}{{\tt arXiv:2503.11293}}.

\bibitem{Andriot:2025cyi}
D.~Andriot, N.~Cribiori, and T.~Van~Riet, {\it {Scale separation, rolling solutions and entropy bounds}},  \href{http://arxiv.org/abs/2504.08634}{{\tt arXiv:2504.08634}}.

\bibitem{Ghoshal:2025tlk}
A.~Ghoshal, F.~Revello, and G.~Villa, {\it {Cosmic superstrings in large volume compactifications: PTAs, LISA and time-varying tension}},  \href{http://arxiv.org/abs/2504.20994}{{\tt arXiv:2504.20994}}.

\bibitem{Conlon:2008cj}
J.~P. Conlon, R.~Kallosh, A.~D. Linde, and F.~Quevedo, {\it {Volume Modulus Inflation and the Gravitino Mass Problem}},  {\em JCAP} {\bf 09} (2008) 011, [\href{http://arxiv.org/abs/0806.0809}{{\tt arXiv:0806.0809}}].

\bibitem{Cicoli:2015wja}
M.~Cicoli, F.~Muia, and F.~G. Pedro, {\it {Microscopic Origin of Volume Modulus Inflation}},  {\em JCAP} {\bf 12} (2015) 040, [\href{http://arxiv.org/abs/1509.07748}{{\tt arXiv:1509.07748}}].

\bibitem{Aldazabal:2000sa}
G.~Aldazabal, L.~E. Ibanez, F.~Quevedo, and A.~M. Uranga, {\it {D-branes at singularities: A Bottom up approach to the string embedding of the standard model}},  {\em JHEP} {\bf 08} (2000) 002, [\href{http://arxiv.org/abs/hep-th/0005067}{{\tt hep-th/0005067}}].

\bibitem{Berenstein:2001nk}
D.~Berenstein, V.~Jejjala, and R.~G. Leigh, {\it {The Standard model on a D-brane}},  {\em Phys. Rev. Lett.} {\bf 88} (2002) 071602, [\href{http://arxiv.org/abs/hep-ph/0105042}{{\tt hep-ph/0105042}}].

\bibitem{Verlinde:2005jr}
H.~Verlinde and M.~Wijnholt, {\it {Building the standard model on a D3-brane}},  {\em JHEP} {\bf 01} (2007) 106, [\href{http://arxiv.org/abs/hep-th/0508089}{{\tt hep-th/0508089}}].

\bibitem{Beasley:2008dc}
C.~Beasley, J.~J. Heckman, and C.~Vafa, {\it {GUTs and Exceptional Branes in F-theory - I}},  {\em JHEP} {\bf 01} (2009) 058, [\href{http://arxiv.org/abs/0802.3391}{{\tt arXiv:0802.3391}}].

\bibitem{Conlon:2008wa}
J.~P. Conlon, A.~Maharana, and F.~Quevedo, {\it {Towards Realistic String Vacua}},  {\em JHEP} {\bf 05} (2009) 109, [\href{http://arxiv.org/abs/0810.5660}{{\tt arXiv:0810.5660}}].

\bibitem{Krippendorf:2010hj}
S.~Krippendorf, M.~J. Dolan, A.~Maharana, and F.~Quevedo, {\it {D-branes at Toric Singularities: Model Building, Yukawa Couplings and Flavour Physics}},  {\em JHEP} {\bf 06} (2010) 092, [\href{http://arxiv.org/abs/1002.1790}{{\tt arXiv:1002.1790}}].

\bibitem{Cicoli:2021dhg}
M.~Cicoli, I.~n.~G. Etxebarria, F.~Quevedo, A.~Schachner, P.~Shukla, and R.~Valandro, {\it {The Standard Model quiver in de Sitter string compactifications}},  {\em JHEP} {\bf 08} (2021) 109, [\href{http://arxiv.org/abs/2106.11964}{{\tt arXiv:2106.11964}}].

\bibitem{Quevedo:2014xia}
F.~Quevedo, {\it {Local String Models and Moduli Stabilisation}},  {\em Mod. Phys. Lett. A} {\bf 30} (2015), no.~07 1530004, [\href{http://arxiv.org/abs/1404.5151}{{\tt arXiv:1404.5151}}].

\bibitem{Marchesano:2024gul}
F.~Marchesano, G.~Shiu, and T.~Weigand, {\it {The Standard Model from String Theory: What Have We Learned?}},  {\em Ann. Rev. Nucl. Part. Sci.} {\bf 74} (2024) 113--140, [\href{http://arxiv.org/abs/2401.01939}{{\tt arXiv:2401.01939}}].

\bibitem{Balasubramanian:2005zx}
V.~Balasubramanian, P.~Berglund, J.~P. Conlon, and F.~Quevedo, {\it {Systematics of moduli stabilisation in Calabi-Yau flux compactifications}},  {\em JHEP} {\bf 03} (2005) 007, [\href{http://arxiv.org/abs/hep-th/0502058}{{\tt hep-th/0502058}}].

\bibitem{Conlon:2005ki}
J.~P. Conlon, F.~Quevedo, and K.~Suruliz, {\it {Large-volume flux compactifications: Moduli spectrum and D3/D7 soft supersymmetry breaking}},  {\em JHEP} {\bf 08} (2005) 007, [\href{http://arxiv.org/abs/hep-th/0505076}{{\tt hep-th/0505076}}].

\bibitem{Brustein:1992nk}
R.~Brustein and P.~J. Steinhardt, {\it {Challenges for superstring cosmology}},  {\em Phys. Lett. B} {\bf 302} (1993) 196--201, [\href{http://arxiv.org/abs/hep-th/9212049}{{\tt hep-th/9212049}}].

\bibitem{Copeland:2003bj}
E.~J. Copeland, R.~C. Myers, and J.~Polchinski, {\it {Cosmic F and D strings}},  {\em JHEP} {\bf 06} (2004) 013, [\href{http://arxiv.org/abs/hep-th/0312067}{{\tt hep-th/0312067}}].

\bibitem{Vilenkin:2000jqa}
A.~Vilenkin and E.~P.~S. Shellard, {\em {Cosmic Strings and Other Topological Defects}}.
\newblock Cambridge University Press, 7, 2000.

\bibitem{Copeland:2009ga}
E.~J. Copeland and T.~W.~B. Kibble, {\it {Cosmic Strings and Superstrings}},  {\em Proc. Roy. Soc. Lond. A} {\bf 466} (2010) 623--657, [\href{http://arxiv.org/abs/0911.1345}{{\tt arXiv:0911.1345}}].

\bibitem{Yamaguchi:2005gp}
M.~Yamaguchi, {\it {Cosmological evolution of cosmic strings with time dependent tension}},  {\em Phys. Rev. D} {\bf 72} (2005) 043533, [\href{http://arxiv.org/abs/hep-ph/0503227}{{\tt hep-ph/0503227}}].

\bibitem{Ichikawa:2006rw}
K.~Ichikawa, T.~Takahashi, and M.~Yamaguchi, {\it {Implications of cosmic strings with time-varying tension on CMB and large scale structure}},  {\em Phys. Rev. D} {\bf 74} (2006) 063526, [\href{http://arxiv.org/abs/hep-ph/0606287}{{\tt hep-ph/0606287}}].

\bibitem{Cheng:2008ma}
H.-b. Cheng and Y.-q. Liu, {\it {The Circular loop equation of a cosmic string with time-varying tension}},  {\em Mod. Phys. Lett. A} {\bf 23} (2008) 3023--3030, [\href{http://arxiv.org/abs/0801.2808}{{\tt arXiv:0801.2808}}].

\bibitem{Sadeghi:2009wx}
J.~Sadeghi, H.~M. Farahani, B.~Pourhassan, and S.~M. Noorbakhsh, {\it {Cosmic string in the BTZ Black Hole background with time-dependant tension}},  {\em Phys. Lett. B} {\bf 703} (2011) 14--19, [\href{http://arxiv.org/abs/0903.0292}{{\tt arXiv:0903.0292}}].

\bibitem{Wang:2012naa}
L.-L. Wang and H.-B. Cheng, {\it {The evolution of circular loops of a cosmic string with periodic tension}},  {\em Phys. Lett. B} {\bf 713} (2012) 59--62, [\href{http://arxiv.org/abs/1206.2095}{{\tt arXiv:1206.2095}}].

\bibitem{Emond:2021vts}
W.~T. Emond, S.~Ramazanov, and R.~Samanta, {\it {Gravitational waves from melting cosmic strings}},  {\em JCAP} {\bf 01} (2022), no.~01 057, [\href{http://arxiv.org/abs/2108.05377}{{\tt arXiv:2108.05377}}].

\bibitem{Amendola:1999qq}
L.~Amendola, {\it {Scaling solutions in general nonminimal coupling theories}},  {\em Phys. Rev. D} {\bf 60} (1999) 043501, [\href{http://arxiv.org/abs/astro-ph/9904120}{{\tt astro-ph/9904120}}].

\bibitem{Amendola:1999er}
L.~Amendola, {\it {Coupled quintessence}},  {\em Phys. Rev. D} {\bf 62} (2000) 043511, [\href{http://arxiv.org/abs/astro-ph/9908023}{{\tt astro-ph/9908023}}].

\bibitem{Copeland:1997et}
E.~J. Copeland, A.~R. Liddle, and D.~Wands, {\it {Exponential potentials and cosmological scaling solutions}},  {\em Phys. Rev. D} {\bf 57} (1998) 4686--4690, [\href{http://arxiv.org/abs/gr-qc/9711068}{{\tt gr-qc/9711068}}].

\bibitem{wainwright1997dynamical}
J.~Wainwright and G.~F.~R. Ellis, eds., {\em Dynamical Systems in Cosmology}.
\newblock Cambridge University Press, Cambridge, 1997.

\bibitem{Bahamonde:2017ize}
S.~Bahamonde, C.~G. B{\"o}hmer, S.~Carloni, E.~J. Copeland, W.~Fang, and N.~Tamanini, {\it {Dynamical systems applied to cosmology: dark energy and modified gravity}},  {\em Phys. Rept.} {\bf 775-777} (2018) 1--122, [\href{http://arxiv.org/abs/1712.03107}{{\tt arXiv:1712.03107}}].

\bibitem{Frey:2023khe}
A.~R. Frey, R.~Mahanta, A.~Maharana, F.~Muia, F.~Quevedo, and G.~Villa, {\it {String thermodynamics in and out of equilibrium: Boltzmann equations and random walks}},  {\em JHEP} {\bf 03} (2024) 112, [\href{http://arxiv.org/abs/2310.11494}{{\tt arXiv:2310.11494}}].

\bibitem{Frey:2024jqy}
A.~R. Frey, R.~Mahanta, A.~Maharana, F.~Quevedo, and G.~Villa, {\it {Gravitational waves from high temperature strings}},  {\em JHEP} {\bf 12} (2024) 174, [\href{http://arxiv.org/abs/2408.13803}{{\tt arXiv:2408.13803}}].

\bibitem{Mitchell:1987hr}
D.~Mitchell and N.~Turok, {\it {Statistical Mechanics of Cosmic Strings}},  {\em Phys. Rev. Lett.} {\bf 58} (1987) 1577.

\bibitem{Mitchell:1987th}
D.~Mitchell and N.~Turok, {\it {Statistical Properties of Cosmic Strings}},  {\em Nucl. Phys. B} {\bf 294} (1987) 1138--1163.

\bibitem{Atick:1988si}
J.~J. Atick and E.~Witten, {\it {The Hagedorn Transition and the Number of Degrees of Freedom of String Theory}},  {\em Nucl. Phys. B} {\bf 310} (1988) 291--334.

\end{thebibliography}\endgroup
\end{document}